\documentclass[aps,pra,showpacs,twocolumn]{revtex4}
\usepackage{bm}
\usepackage{amsmath}
\usepackage{amssymb}
\usepackage{graphicx}
\usepackage{epsfig}
\usepackage{epstopdf}
\begin{document}

\def\pr{\prime}
\def\be{\begin{equation}}
\def\en#1{\label{#1}\end{equation}}
\def\d{\dagger}
\def\bar#1{\overline #1}
\newcommand{\per}{\mathrm{per}}

\newcommand{\rd}{\mathrm{d}}
\newcommand{\vare}{\varepsilon }
\newcommand{\m}{\mathbf{m}}
\newcommand{\bl}{{\boldsymbol\ell}}
\newcommand{\bt}{\mathbf{t}}
\newcommand{\cJ}{\mathcal{J}}

\title{   Distinguishability theory for   time-resolved photodetection  and   boson sampling     }

\author{V. S. Shchesnovich and M. E. O. Bezerra}

\address{Centro de Ci\^encias Naturais e Humanas, Universidade Federal do
ABC, Santo Andr\'e,  SP, 09210-170 Brazil }

\begin{abstract}

We   study distinguishability   of photons in multiphoton interference on a multiport  when   fast detectors, capable of precise time resolution, are employed.  Such a  setup  was previously suggested for experimental realization of boson sampling with single  photons.   We  investigate if  fast  photodetection   allows  to circumvent   distinguishability of    realistic  single photons in mixed  states. To this goal we   compare    distinguishability  of photons     in  two setups:  (a)   with photons in  the same  average (temporal) profile  on a spatial interferometer and  photodetection incapable of (or with strongly imprecise) time resolution  and (b)  with  photons in generally  different  average   temporal  profiles  on the same spatial interferometer  and  photodetection  with precise time  resolution.  Exact  analytical results are obtained for  Gaussian-shaped   single photons      with Gaussian   distribution of  photon arrival  time.  Distinguishability  of photons in the  two setups  is found to be strikingly  similar.  For the same purity of photon states,  only the same   quality experimental boson sampling can be achieved using either of   the two   setups.   The upshot of our results   is that    distinguishability   due to mixed states   is an intrinsic property of  photons,  whatever the photodetection scheme. 
\end{abstract}

\maketitle

\section{Introduction}
\label{sec1}

We have witnessed  enormous experimental progress  \cite{Reviewbs} in experimental realization of    boson sampling    \cite{AA}  for  demonstration of quantum advantage   over  classical computations.  Recent experiment with   20  input photons on a  60-mode interferometer  \cite{20ph60mod}  is a big step towards the ultimate    goal of  demonstrating the quantum advantage.  For such a    quantum device, scaled up beyond possibility to exhaust all outputs in an experiment,   it is of paramount importance to establish to what extent the   quantum advantage     survives  sources of noise/imperfections  of an experimental setup and whether one could  circumvent at least some  imperfections  in some way. One  important source of noise/imperfection is, of course, photon distinguishability \cite{HOM}, whos effect is   exponential  with  number of photons       \cite{R1,VS2014,TightBound} allowing at certain scale for efficient    simulation of    experimental boson sampling  on classical computers \cite{Jelmer2018}.  
 Besides distinguishability of photons, there are many other   sources of noise/imperfections  affecting  optical boson sampling  devices, such as  noise in  interferometer \cite{KK,LP,A},  photon losses and     random counts of detectors \cite{AB,PRS,OB}.   The  quality of an experimental boson sampling device  will depend on all such sources of noise, recently    shown  to satisfy     equivalence relations  \cite{VS2019}.

With respect to  distinguishability of photons   in a large-scale multiphoton interference experiment,  such as  boson sampling with single photons,  notwithstanding    many  theoretical and experimental results \cite{VS2014,MPBF,SUN,Rohde,PartDist,Tichy,TL1,TL,MCMS,VSMB,DD,QSPD,CrSp,Exp1,Exp2},  there  still remains  an issue lacking a  detailed  investigation. Namely,    to what extent it is possible to circumvent the distinguishability  by using  photodetection capable of resolution of  internal states of photons,  for instance,   by photodetection with precise resolution in time (i.e., by using fast  detectors  as compared to  photon   pulse  duration).  It was previously   suggested \cite{TL1,TL}  that   such time-resolved photodetection can compensate  distinguishability of photons, though  the result   is limited  to  photons in pure states only.  
     
In this work we  investigate  the above issue by extending   the theory of    partial distinguishability    \cite{VS2014,PartDist}  to  the case of     photodetection with  precise    time  resolution. We consider  multiphoton interference in general  with  application of the results   to boson sampling with single photons.  Here we note that the effect of internal state discrimination at the detection stage  was briefly considered in Ref. \cite{VSMB}, without  full analysis of the effect on  distinguishability, whereas  Ref. \cite{DD}  analyzed mixed-state distinguishability within an approach based on symmetry expansion of multiphoton state, not equivalent to that adopted below. 
In doing this, we account for   \textit{realistic}   single photons  in   mixed states, in contrast to related  works     \cite{TL1,TL} limited only to  \textit{pure-state} photons.      Pure-state photons, however,  are  not  available in     experiments  \cite{SPnp} due to unavoidable sources of  noise.  For instance,   the standard spontaneous parametric downconversion sources of single photons  \cite{SPDCtjitt,HervsPurity}  as well as the     quantum dot sources      \cite{QDtjitt} lead to  random emission times (for instance, in the former case    only  single photons in  ``thermal-difference"  mixed state can be  achieved  \cite{HeraldSP}).  Additionally,   realistic detectors act   by projection  onto mixed states \cite{Rdetec1,Rdetec2}, with probabilities of   outcomes given  by positive-operator valued measures.  We use  mixed-state  single photons  with  random arrival times to  model such  sources of noise in experiments.     
 Single photons with  near-unity purity/indistinguishability have  been recently      experimentally  demonstrated      \cite{SP1,SP2}.   It remains to  establish  if  the achieved  purity  is good enough  when scaling up to   a large number   of photons  \cite{QSBS},   as required for quantum advantage with boson sampling.

 To better understand the  effect of  photon  distinguishability due to mixed states    in  case of   photodetection with precise time   resolution, we  compare with the  well-studied case of photodetection    incapable of (or with  strongly imprecise)   time resolution,  first  considered in Ref.  \cite{VS2014}.    We therefore compare  distinguishability of photons in  two setups   of  multiphoton interference:  (a)   with single photons of  the same  central frequency   on a spatial interferometer with  photodetection   incapable of time  resolution (slow  detectors  as compared to  pulse duration of photons)  and   (b)  with single photons of generally different   central frequencies on the same spatial interferometer    and   photodetection with precise     time resolution (fast detectors). Our approach is applicable to any mixed states of photons,  however,  for  explicit analytical results we  use     Gaussian-shaped photons with Gaussian distributed photon arrival  times.  
 
 We also generalize  previously considered  measure   \cite{TightBound}  of  quality of photon indistinguishability    to  the   case of photodetection with precise time resolution. Our measure  coincides  with the total  probability that bosons behave as completely indistinguishable particles  in multiphoton  interference on a  multiport.   It turns out that the two considered setups  correspond to  \textit{the same} value of this measure of indistinguishability.

   The  upshot   of our work  is that  the  (overall) distinguishability due to impure (mixed) states of photons cannot be circumvented by  any photodetection scheme, even if it is capable of precise photon  state resolution.    This means, for instance,  that    using photons of  different frequencies and   photodetection  with  strongly precise time resolution, as compared to photon pulse duration,  will not  allow for a better quality  (i.e., closer to   ideal)     boson sampling in comparison with photons of the same frequency and photodetection with strongly imprecise time resolution.  

The text is organized as follows. In section \ref{sec2}  we introduce our model of Gaussian single photons,  briefly recall   known facts on  distinguishability of photons, directly applicable to our  setup (a), with  photodetection incapable of time resolution,   subsection \ref{sec2A}, and    then discuss  the effect of photon distinguishability in  setup (b), with  photodetection capable of  precise  time resolution, subsection \ref{sec2B}.      In section \ref{sec3} we  generalize   to  setup (b)  a measure of  quality of photon indistinguishability, introduced previously   for setup (a),  and compare the quality of approximation to the ideal boson sampling by the two setups. In concluding section \ref{sec4} we discuss  implications of our results. Some mathematical details of derivations are relegated to appendices \ref{appA} and \ref{appB}.

\section{Two setups of multiphoton interference:   with and without photon arrival time resolution by detectors   }
\label{sec2}

\begin{figure}[htb]
\begin{center}
     \includegraphics[width=.45\textwidth]{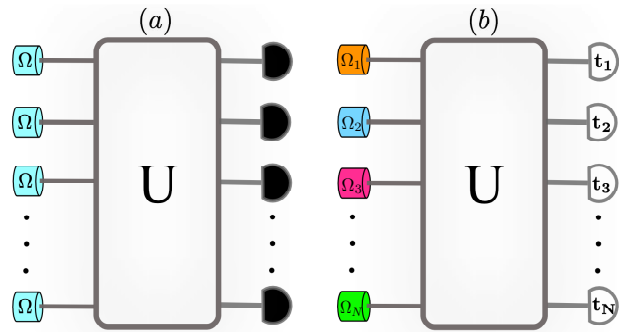}
     \caption{A schematic depiction of our two setups  with single photons: (a) $N$  photons of the same central frequency   $\Omega$  interfere on a unitary  spatial $M$-port interferometer  $U$ and are detected at  output ports without photodetection times resolution (slow  detectors), with the output data being  $l_1,\ldots,l_N$ (order is irrelevant); (b) $N$ photons of   frequencies    $\Omega_1,\Omega_2,\ldots, \Omega_N$   interfere on the same spatial interferometer $U$ and are detected with  precise resolution of   photodetection times, with the output data being $(l_1,t_1), \ldots, (l_N,t_N)$.  In general $N\le M$ and $l_k= l_j$ and/or $t_k= t_j$ for some $k\ne j$.    \label{F1} }
   \end{center}
\end{figure}

We consider two  setups of multiphoton interference  with single photons on the same spatial interferometer, with and without photon arrival time resolution by  detectors,   respectively, panels (a) and (b) of figure \ref{F1}, where in setup (a) single photons have  the same central frequency, whereas  in setup (b) they  have, in general,  different central frequencies.  We assume that a unitary linear  spatial interferometer,  with $M$ input and output ports,  has all   paths of equal optical length (independent of the frequency range of photons). Such an  interferometer  can be  described by a  unitary  transformation   between  the   $M$  input  and output spatial  modes (ports)
\be
 \hat{a}^\dag_k(t) = \sum_{l=1}^M U_{kl}  \hat{b}^\dag_l(t),
\en{Eq2}
where $U$ is a unitary matrix, and  $\hat{a}_k(t)$ and $\hat{b}_l(t)$ are boson creation operators of the input and output ports at time $t$.    Below we will  always  employ index $k$  for the input port of an interferometer (or for the photon originating from  input port $k$), whereas    the output ports  will be labelled by $l$.  We will also use $m_l$ for  the  number of photons detected at output port $l$ and denote   $\m=(m_1,\ldots,m_M)$,  $m_1+\ldots + m_M=N$,   an output  configuration of  $N$ detected photons. Note that in general,     $l_1,\ldots, l_N$ is a multi-set   of output ports (i.e., containing repetitions), where the order is not significant, since  the   probability of detecting photons is    symmetric in $l_1,\ldots, l_N$, being dependent only on $\m$.  Let  us fix the   input  ports  of single photons  to be $k=1,\ldots, N$.

Our principal results and  conclusions  apply for general  (mixed)  states  $\hat{\rho}_1,\ldots,\hat{\rho}_N$ of input   photons, however,  we   illustrate the approach by utilizing  the simplest  Gaussian model, where each photon has a Gaussian shape and   photon  arrival times are distributed   according to a Gaussian. Besides allowing explicit analytical results, the   Gaussian-shaped single photons  are optimal for multiphoton interference experiments  \cite{Gphotons}.  Thus, we consider $ N$ single photons  in the following mixed  states:
\begin{eqnarray}
\label{Eq1}
&& \hat{\rho}_k = \int \rd \tau p_k(\tau)|\Phi_{k,\tau}\rangle\langle \Phi_{k,\tau}|,\quad p_k(\tau) = \frac{1}{\sqrt{\pi}\Delta\tau_k} e^{-\frac{\tau^2}{\Delta\tau_k^2}} \nonumber\\
&& |\Phi_{k,\tau}\rangle  =  \int\rd t\, \Phi_{k,\tau}(t)\hat{a}^\dag_k(t)|0\rangle \nonumber\\
&&  \Phi_{k,\tau}(t) = \frac{1}{\pi^{1/4}\sqrt{T_k}}\exp\left(-i\Omega_k t - \frac{(t- \tau)^2}{2T_k^2}\right),
\end{eqnarray}
where $\Delta\tau_k$  is  the standard deviation of arrival time of photon   $k$, $T_k$ is the  temporal width of the photon pulse, and  $\Omega_k$ is  its  central frequency.   In Eq. (\ref{Eq1}) we have assumed that   the Gaussian distributed  arrival  times of photons have the same average  ($\overline{\tau} =0$), since our main objective is to study the effect of randomness (in photon arrival times)  in the two setups of figure \ref{F1}.  On the other hand,  constant bias in the  average arrival times of  single photons can be  compensated for  with  delay lines. Photons in   panels (a) and (b) of figure \ref{F1}  differ in  central frequency in  Eq. (\ref{Eq1}): in setup (a)   photons have the same central frequency $\Omega_k = \Omega$,  whereas in setup (b)   central frequencies are  generally  different,   $\Omega_k\ne \Omega_l$ for $k\ne l$.

\subsection{Setup (a):  no time resolution by detectors  } 
\label{sec2A}

Here we briefly recall the partial distinguishability theory \cite{VS2014,PartDist} directly applicable to setup (a) of figure \ref{F1}.  We therefore  set the same  central frequency $\Omega_k=\Omega$ in Eq. (\ref{Eq1}).    An explicit    expression for  the   distinguishability function of our  Gaussian   model and an exponential approximation (see Eqs. (\ref{Eq7})-(\ref{Eq8}) and (\ref{JA}) below)     are nevertheless  new results.    Photodetection without arrival time resolution  is described by the  following positive operator valued measure (POVM) \cite{VS2014} 
\be
\hat{\Pi}_\m = \frac{1}{\m!} \int\rd t_1\ldots \int\rd t_N\left[ \prod_{k=1}^N\hat{b}^\dag_{l_k}(t_k)\right] |0\rangle\langle 0| \prod_{k=1}^N \hat{b}_{l_k}(t_k),
\en{Eq3}
where $\m=(m_1,\ldots,m_M)$,    $m_l$ is the  number of photons in output port  $l$,   and $\m! = m_1!\ldots m_M!$.  
 The probability $\tilde{p}_\m$  of a given  output configuration $\m$ is  
 \be 
\tilde{p}_\m = \mathrm{Tr}\{\hat{\Pi}_\m \hat{\rho}_1\otimes\ldots \otimes  \hat{\rho}_N\}.  
\en{pm}
To compute the probability, we observe that Eq. (\ref{Eq2}) leads to the following   identity  \cite{VS2014} 
\be
\langle 0|\!\!\left[\prod_{k=1}^N\hat{b}_{l_k}(t_k)\right] \prod_{k=1}^N\hat{a}^\dag_k(t^\prime_k)|0\rangle = \sum_\sigma \prod_{k=1}^N\! U_{\sigma(k),l_k}\delta(t^\prime_k-t_{\sigma^{-1}(k)}),
\en{id1}
where $\sigma$ is a permutation of $N$ objects. Identity (\ref{id1})   allows  straightforward  computation of the inner products involving boson operators in Eq. (\ref{pm}).  In particular,  for   the integrand  $\prod\limits_{k=1}^N |\Phi_{k,\tau_k}\rangle\langle\Phi_{k,\tau_k}|$ in  the   integral over $\tau_1,\ldots, \tau_N$  in the  input state   $\hat{\rho}_1\otimes \ldots \otimes \hat{\rho}_N$, with $\hat{\rho}_k$ of  Eq. (\ref{Eq1}), we get   the following inner product (and a similar   conjugated one)  
 \begin{eqnarray}
 \label{id2}
 && \int\rd t^\prime_1 \ldots \int\rd t^\prime_N\langle 0|\left[\prod_{k=1}^N \hat{b}_{l_k}(t_k)\right] \prod_{k=1}^N \Phi_{k,\tau_k}(t^\prime_k)\hat{a}^\dag_{k}(t^\prime_k)  |0\rangle \nonumber\\
 &&= \sum_\sigma \prod_{k=1}^N  U_{\sigma(k),l_k} \Phi_{k,\tau_k}(t_{\sigma^{-1}(k)}). 
 \end{eqnarray}
The probability $\tilde{p}_\m$   becomes \cite{VS2014,PartDist}
\be   
\tilde{p}_\m = \frac{1}{\m!}\sum_{\sigma_1,\sigma_2} J(\sigma_1\sigma^{-1}_2) \prod_{k=1}^N U^*_{\sigma_1(k),l_k} U_{\sigma_2(k),l_k},
\en{Eq4}
with a distinguishability function $J(\sigma)$, given in this case by 
\be
J(\sigma)  =  \mathrm{Tr}\left(P^\dag_\sigma \rho_1 \otimes \ldots \otimes \rho_N \right) ,
\en{Eq5}
where we have used the unitary  operator representation $P_\sigma$ ($P^\dag_\sigma = P_{\sigma^{-1}}$)  of  permutation $\sigma$, defined by 
\be
P_\sigma |x_1\rangle\otimes\ldots \otimes|x_N\rangle =  |x_{\sigma^{-1}(1)}\rangle\otimes\ldots \otimes|x_{\sigma^{-1}(N)}\rangle,
\en{Psigma}
\\[0.005cm]
and introduced the internal state $\rho_k$ of photon  $k$, defined as follows  
\begin{eqnarray} 
\label{Eq6}
 && \rho_k  \equiv \int \rd \tau p_k(\tau)|\phi_{k,\tau}\rangle\langle \phi_{k,\tau}|, \quad 
  |\phi_{k,\tau}\rangle  =  \int\rd t\, \phi_{k,\tau}(t) |t\rangle,\nonumber\\
  && \phi_{k,\tau}(t) = \frac{1}{\pi^{1/4}\sqrt{T_k}}\exp\left( - \frac{(t- \tau)^2}{2T_k^2}\right),
 \end{eqnarray}
 where $|t\rangle$ is the time-basis state $\langle t|t^\prime\rangle = \delta(t-t^\prime)$ and   $p_k(\tau)$ is given by Eq. (\ref{Eq1}) (note  that    $\Phi_{k,\tau}(t) = e^{-i\Omega_k t}\phi_{k,\tau}(t)$).  
 
 \subsubsection*{Identical mixed  internal states: $\rho_k = \rho$}
 
 Distinguishability due to mixed internal states, e.g. as in  Eq. (\ref{Eq6}), is similar to the usual pure-state-overlap distinguishability \cite{HOM},  e.g.,  for two photons with internal states  $\rho_1$ and $\rho_2$ the   average absolute value squared of  the state overlap    is     $\mathrm{Tr}(\rho_1\rho_2)$. However, differently from the pure-state distinguishability, taking two photons in the same  mixed internal state $\rho_{1,2}=\rho$ does not make them indistinguishable, since  $\mathrm{Tr}(\rho^2)\ne 1$. 
  Therefore, the effect of distinguishability is already witnessed  in the simplest case, when the internal mixed states are the same, which we consider in what follows.  In calculations,  we will use 
 $\Delta\tau_k=\Delta\tau$ and $T_k=T$   (see Eqs. (\ref{Eq1}) and (\ref{Eq6})), so that $ \rho_k= \rho$, for all $k=1,\ldots, N$.   In this case the expression for distinguishability function $J(\sigma)$ of Eq. (\ref{Eq5}) simplifies considerably \cite{VS2014,PartDist}:
 \be
 J(\sigma) = \prod_{n=2}^N \mathrm{Tr}\left(\rho^n\right)^{C_n(\sigma)},
 \en{Eq7}
 where $C_n(\sigma)$ is the number of permutation cycles  of length $n$  in the disjoint cycle decomposition of $\sigma$, i.e., cycles of the type  $i_1\to i_2\to \ldots \to i_n\to i_1$ (for more information  see Ref. \cite{Stanley}). The higher-order purities $ \mathrm{Tr}\left(\rho^n\right)$ for $n=2,\ldots,N$ govern   partial  distinguishability of $N$ single photons in  the same  (mixed) internal state $\rho$. For the   Gaussian model of Eq. (\ref{Eq6}),  moreover, the purities can be evaluated explicitly, as this amounts to evaluating multidimensional Gaussian integrals. Let us introduce   the relative uncertainty in  time of arrival by  dividing  the standard deviation of photon arrival  time by its  pulse duration  $\eta \equiv  \frac{\Delta\tau }{2T } $. From Eqs. (\ref{Eq5}) and (\ref{Eq6})   we obtain  for $n\ge 2$  (for more details, see appendix \ref{appA})
  \begin{eqnarray}
 \label{Eq8}
&&  \mathrm{Tr}\left(\rho^n\right) = (1+2\eta^2)^{-\frac{n}{2}} (X^n_+ - X^n_-)^{-1},  \nonumber\\
  && X^2_\pm = \frac12\left(1\pm \frac{\sqrt{1+4\eta^2}}{1+2\eta^2} \right).  
\end{eqnarray}
When $\eta\to 0$  we get  $ \mathrm{Tr}\left(\rho^n\right) =1$ and thus $J(\sigma)=1$, i.e., the ideal case with  completely indistinguishable bosons. For    small $\eta\ll1$, by  expanding $\ln(1+2\eta^2) = 2\eta^2+ O(\eta^4)$ and $X_+^n-X_-^n= 1 + O(n\eta^4)$, we arrive at  an exponential approximation 
 \begin{eqnarray}
 \label{Eq8A}
  \mathrm{Tr}\left(\rho^n\right)  =  e^{- n (\eta^2 +O(\eta^4))} \approx e^{-n\eta^2}
 \end{eqnarray}
  (see also Fig. \ref{F2} in appendix \ref{appA} for a numerical comparison of the expressions in Eqs. (\ref{Eq8}) and  (\ref{Eq8A})).
Substituting   approximation    (\ref{Eq8A}) into Eq. (\ref{Eq7})  gives
\be
 J(\sigma)  \approx  \exp\left(- \eta^2 [N-C_1(\sigma)]\right),
\en{JA}
where we have used that  the cycle lengths add up to $N$: $\sum_{n=1}^N n C_n(\sigma) = N$. 

 \subsubsection*{Rescaling the  output probability of setup  (a)  }
 \label{sec2Asub}
 
 Below we will  compare the output probability in two setups of figure  \ref{F1}.   Setup (b)   involves  continuous basis of  states in time (see the next section),  with the usual problem of normalization for continuous bases.    Instead of    introducing  an arbitrary countable state basis \cite{ContFields} for setup (b),   it is   more convenient   to use   the unnormalized symmetric basis states. For comparison of the two setups, we have  to rescale the  output probability   of setup (a)  in such a way that similar unnormalized symmetric basis states at the output are used, instead of the usual Fock states.    
 
Let us    rescale the output  probability of  Eq.  (\ref{Eq4})  as follows   $p_\bl \equiv \frac{\m!}{N!}\tilde{p}_\m$, where  a multi-set  of output ports $\bl = (l_1,\ldots, l_N)$  corresponds to   output   configuration  $\m$. Observe that, due to the summation identity for any symmetric function $f(l_1,\ldots,l_N)$
\be
\sum_\m f(l_1,\ldots,l_N) =  \sum_{l_1=1}^M \ldots \sum_{l_N=1}^M  \frac{\m!}{N!} f(l_1,\ldots,l_N),
\en{idSUM}
 the normalization of probability   $\sum_\m \tilde{p}_\m =1$ becomes 
   \[
 \sum_{l_1=1}^M \ldots \sum_{l_N=1}^M  p_{\bl} = 1,
 \]
 i.e.,  instead of occupation numbers $\m$ (output configuration) the multi-set of output ports $l_1,\ldots,l_N$  is \textit{formally} considered as  the output data. From now on,  we will also employ  the  superscript, $(a)$ or   $(b)$, to distinguish the   probability distributions of the two setups in figures \ref{F1}(a) and \ref{F1}(b). We obtain for the setup in figure \ref{F1}(a)
 \begin{eqnarray}
\label{Eq10}
 p^{(a)}_{\bl} & = & \mathrm{Tr}\{\Pi_{\bl} \hat{\rho}_1\otimes\ldots \otimes  \hat{\rho}_N  \} \nonumber\\
 &= &\frac{1}{N!}\sum_{\sigma_1,\sigma_2} J(\sigma_1\sigma_2^{-1} ) \prod_{k=1}^N U^*_{\sigma_1(k),l_k} U_{\sigma_2(k),l_k},
\end{eqnarray}
  with the  POVM  (compare with the Fock-state POVM $\hat{\Pi}_\m$ of Eq. (\ref{Eq3}))
 \be
 \Pi_{\bl} =  \frac{1}{N!} \int\rd t_1\ldots \int\rd t_N \left[\prod_{k=1}^N\hat{b}^\dag_{l_k}(t_k)\right] |0\rangle\langle 0| \prod_{k=1}^N \hat{b}_{l_k}(t_k),
 \en{Eq11} 
where $\frac{1}{\sqrt{N!}}\prod_{k=1}^N\hat{b}_{l_k}(t_k) |0\rangle$ is the mentioned above unnormalized symmetric basis state of $N$ bosons. The POVM of Eq. (\ref{Eq11})  is  positive semidefinite and normalized, since we have 
\be
\sum_{l_1=1}^M \ldots \sum_{l_N=1}^M \Pi_{\bl} =  \frac{1}{N!}  \sum_\sigma P_\sigma  \equiv S_N,
 \en{Eq12}
 where $ S_N$ is the projector on the symmetric states of $N$ particles, i.e.,  the identity operator in the Hilbert space of $N$ bosons.

 \subsection{Setup (b):  precise time resolution by detectors} 
\label{sec2B}

Consider now the  setup of figure \ref{F1}(b) where,  instead of introducing the actual detection time $\delta t\ll T$, for simplicity, we  use the  instantaneous detection model, thus  we will consider   probability density at output. The   POVM density $\Pi_\bl(\bt)$, where $\bt= (t_1, \ldots, t_N)$,  describing such sharply precise  photodetection  can be  obtained from that of Eq. (\ref{Eq11}) by removing the integrals:
 \be
 \Pi_\bl(\bt) \equiv \frac{1}{N!}  \left[\prod_{k=1}^N\hat{b}^\dag_{l_k}(t_k)\right] |0\rangle\langle 0| \prod_{k=1}^N \hat{b}_{l_k}(t_k).
 \en{Eq13}
The probability density $p_\bl(\bt)$ of detecting $N$ input photons in the output ports $\bl$  at times $\bt$ is accordingly 
 \be
 p^{(b)}_\bl(\bt) =    \mathrm{Tr}\{\Pi_{\bl}(\bt) \hat{\rho}_1\otimes\ldots \otimes  \hat{\rho}_N  \}.
 \en{Eq14}
 Similar as in the previous section, using   the identity of Eq. (\ref{id1}) one can easily compute the inner products in   Eq. (\ref{Eq14}) involving the boson creation and annihilation operators. By comparing   Eqs. (\ref{Eq14}) and  (\ref{Eq10})  one concludes that   computation  of the above inner products  amounts to repeating  that of    section \ref{sec2A}, however with the   integration over   photodetection times  removed, and  rescaling.  The following result is obtained  \begin{eqnarray}
 \label{Eq15}
&& \!\!\!\!  p^{(b)}_\bl(\bt)\! = \!  \frac{1}{N!} \sum_{\sigma_1,\sigma_2}\cJ(\bt;\sigma_1,\sigma_2) \prod_{k=1}^N \mathcal{U}^*_{\sigma_1(k),l_k}\!(t_k) \mathcal{U}_{\sigma_2(k),l_k}\!(t_k), \nonumber\\
 && \!\!\!\! \mathcal{U}_{k,l}(t)  \equiv e^{-i\Omega_k t} U_{k,l},
 \end{eqnarray}
 with the  distinguishability function given  as follows (compare with Eq. (\ref{Eq5}))
 \begin{eqnarray}
 \label{Eq16}
 && \!\!\!\cJ(\bt;\sigma_1,\sigma_2) \! = \!\prod_{k=1}^N \!\int\rd \tau_k p_k(\tau_k)   \phi^*_{k,\tau_k}(t_{\sigma^{-1}_1(k)})\phi_{k,\tau_k}( t_{\sigma^{-1}_2(k)}) \nonumber\\
&  & =  \mathrm{Tr}\left( \!P^\dag_{\sigma_1} |t_1\rangle\langle t_1| \otimes \ldots \otimes |t_N\rangle\langle t_N|P_{\sigma_2} \rho_1\otimes \ldots\otimes \rho_N \right),\nonumber\\
&& = \prod_{k=1}^N \langle t_{\sigma_2(k)}|\rho_k|t_{\sigma_1(k)}\rangle 
  \end{eqnarray}
 where  the inner state $\rho_k$  of photon $k$ is given in  Eq. (\ref{Eq6}) and    $P_\sigma$ is defined in Eq. (\ref{Psigma}).  
 
 Note that the internal states of photons that  define  the  distinguishability functions in Eqs. (\ref{Eq5}) and (\ref{Eq16})  are the same and given by Eq. (\ref{Eq6}),  i.e., irrespectively, whether we consider the setup with or without  precise time resolution. 
  Moreover, the following  relation holds between the distinguishability functions of the two setups 
 \be
 \int\rd t_1\ldots \int \rd t_N \cJ(\bt;\sigma_1,\sigma_2) = J(\sigma^{-1}_1\sigma_2).
 \en{Eq17}
 
Distinguishability function in the form similar to that of Eq. (\ref{Eq16}) has  appeared before in discussion of the effect of  state resolving photodetection on multiphoton interference \cite{VSMB},  however no  analysis of  partial   distinguishability  in such a setup  was  attempted  before.

 For input    photons   in pure states, i.e.,  for   $\Delta\tau_k=0$, the distinguishability function $\cJ$ Eq. (\ref{Eq16})  factorizes as follows 
 \begin{eqnarray}
 \label{Jpure} 
   \cJ(\bt;\sigma_1,\sigma_2) &=& \prod_{k=1}^N  \phi^*_{k,0}(t_{\sigma^{-1}_1(k)})\phi_{k,0}( t_{\sigma^{-1}_2(k)})\nonumber\\
 &= & \prod_{k=1}^N  \phi^*_{\sigma_1(k),0}(t_k)\phi_{\sigma_2(k),0}( t_k)
 \end{eqnarray}
 (in this case  $\tau=0$ in $\phi_{k,\tau}(t)$ of Eq. (\ref{Eq6})).  In this special case, using the relation $\Phi_{k,0}(t) = e^{-i\Omega_k t}\phi_{k,0}(t)$, one   obtains
 \begin{eqnarray}
   p^{(b)}_\bl(\bt)=  \frac{1}{N!} \left| \sum_{\sigma } \prod_{k=1}^N U_{\sigma(k),l_k}\Phi_{\sigma(k),0}( t_k)\right|^2,
      \label{ppure}
 \end{eqnarray}
 i.e., the output probability is given by   absolute value squared of a single matrix permanent of    \mbox{$B_{j,k}\equiv {U}_{j,l_k} \Phi_{j,0}(t_k)$}, where the matrix permanent of an $N$-dimensional matrix $B$ is defined as follows  $\mathrm{per}(B) = \sum_{\sigma}\prod_{k=1}^N B_{\sigma(k),k}$, where the sum is over all permutations $\sigma$ of $N$ objects. 
 Thus, our approach  reproduces the    so-called ``multiboson correlation sampling" with pure-state single photons    \cite{TL}.  In this  case, the photons   are    completely indistinguishable (see also below). However, the   states of photons in any experiment are mixed states due to various sources of noise,  not allowing  the factorization of $\cJ$ as in Eq. (\ref{Jpure}).

When  the distinguishability function in Eq. (\ref{Eq16}) satisfies $\cJ(\bt;\sigma_1,\sigma_2)=0$     whenever $\sigma_1\ne \sigma_2$, there is no multiphoton interference, since the output probability is a convex combination of products of probabilities for each photon. Indeed,   by Eq. (\ref{Eq16})  $0\le \cJ(\bt;\sigma,\sigma) \le 1$,  whereas  from Eq. (\ref{Eq15}) we get
\be
  p^{(b)}_\bl(\bt)=  \frac{1}{N!}  \sum_{\sigma } \cJ(\bt;\sigma,\sigma) \prod_{k=1}^N \left| {\mathcal{U}}_{\sigma(k),l_k}(t_k)\right|^2.
\en{pclass} 
The  output probability density  in Eq. (\ref{pclass})  can be  simulated with classical particles (which are classically indistinguishable). Therefore, it  is  the  classical limit, in accordance with the theory \cite{VS2014,PartDist}. It   occurs   when the overlap 
 between the different photon pulses vanishes, which in the  model of Eq. (\ref{Eq1})  is the limit of  infinite relative uncertainty in arrival times of photons $\Delta\tau_k/T_k\to \infty$.  
 
 \subsubsection*{Identical mixed internal states: $\rho_k=\rho$}

 In general, for different (mixed) internal states of  photons, $\rho_k\ne \rho_j$ for $k\ne j$,  the distinguishability function $\cJ$ of Eq. (\ref{Eq16})  depends on two permutations separately.  
As discussed in the previous section, we focus on the simplest case of photons in the same mixed internal state.   When  $\rho_k = \rho$ in Eq. (\ref{Eq6}) (i.e., $\Delta\tau_k = \Delta\tau$ and $T_k = T$) we get that   $\cJ$ of Eq. (\ref{Eq16}) depends only on the relative permutation, $\cJ(\bt;\sigma_1,\sigma_2) = \cJ(\bt; \sigma_1\sigma^{-1}_2,I)$ ($\sigma=I$ being the identity permutation), thanks   to the following  identities     
 \be
 P_{\sigma_2}  \rho\otimes \ldots \otimes \rho  =    \rho\otimes \ldots \otimes \rho P_{\sigma_2}, \quad P_{\sigma_2}P^\dag_{\sigma_1} = P^\dag_{\sigma_1\sigma^{-1}_2}. 
 \en{id3}
  Below we focus on  this simplified case and  use the notation $\cJ(\bt; \sigma )\equiv \cJ(\bt; \sigma,I)$. Evaluating    Gaussian integrals in Eq. (\ref{Eq16}), we get
 \begin{widetext}
 \be
\cJ(\bt;\sigma) = \frac{1}{\left(\pi[T^2 + \Delta\tau^2]\right)^{N/2}} \exp\left(- \frac{\sum_{k=1}^N t_k^2}{T^2 + \Delta\tau^2} -  \left(\frac{\Delta\tau}{2T}\right)^2 \frac{\sum_{k=1}^N (t_k - t_{\sigma(k)})^2}{T^2 + \Delta\tau^2}  \right).
\en{Eq18}
 \end{widetext}
 Differently from $J(\sigma)$ of the previous section, we have $\cJ(\bt,I)= p(\bt) \ne 1$, where we have introduced   the  probability density  $p(\bt)$ of detecting photons at times $\bt$, irrespective in which  output ports of a spatial interferometer $U$   the input photons are detected. Indeed, by   unitarity of matrix $U$,  we have from Eq. (\ref{Eq15})  (starting with  general $\rho_k$)
 \begin{eqnarray}
 \label{Eq19}
 && p(\bt) \equiv \sum_{l_1=1}^M \ldots \sum_{l_M=1}^M p^{(b)}_\bl(\bt)  \nonumber\\
 && = \frac{1}{N!} \sum_{\sigma_1,\sigma_2}\cJ(\bt;\sigma_1,\sigma_2)  \prod_{k=1}^N \delta_{\sigma_1(k),\sigma_2(k)}e^{i(\Omega_{\sigma_1(k)} - \Omega_{\sigma_2(k)})t_k}\nonumber\\
 &&=  \frac{1}{N!} \sum_{\sigma }\cJ(\bt;\sigma,\sigma) \stackrel{\rho_k \to  \rho}{=} \cJ(\bt,I), 
  \end{eqnarray}  
since  for $\rho_k = \rho$ we have $\cJ(\bt;\sigma,\sigma) = \cJ(\bt;\sigma\sigma^{-1},I) \equiv \cJ(\bt;I) $ and there are exactly $N!$ permutations $\sigma$ of $N$ objects. 

It is   clear   from Eqs. (\ref{pclass}) and (\ref{Eq19}) that  the probability density $p(\bt)$ is also the weight of the respective classical contribution to  output probability density of detecting photons and times $\bt$ (in any output ports $\bl$).  Recall that in the setup  of figure \ref{F1}(a)  the weight of the   classical contribution to output probability is always $J(I)=1$, by the normalization of the distinguishability function,  Eq. (\ref{Eq5}). Therefore,     to    compare  the distinguishability of photons in the two setups,  for each  given set of photodetection  times $\bt$  in    setup  (b),   we rescale $\cJ(\bt;\sigma)$ so that   the classical contribution is also weighted by $1$. We  therefore divide the distinguishability function $\cJ(\bt;\sigma)$ of Eq. (\ref{Eq18}) by the probability density $p(\bt)$ of Eq. (\ref{Eq19}). Hence, the  \textit{proper distinguishability function} of setup (b) becomes   
 \begin{eqnarray}
 \label{Eq20}
 \widetilde{\cJ}(\bt;\sigma) &\equiv & \frac{\cJ(\bt;\sigma)}{p(\bt)}  =  \prod_{k=1}^N  \frac{\langle t_k|\rho|t_{\sigma(k)}\rangle }{   \langle t_k| \rho|t_k\rangle }\nonumber\\
 & = &   \exp\left( -  \eta^2 \frac{\sum_{k=1}^N (t_k - t_{\sigma(k)})^2}{T^2 + \Delta\tau^2}  \right).
 \end{eqnarray}
 
Let us introduce  the following pure states  
  \be
  \chi_k(t)\equiv  \frac{1}{(\pi[T^2 +\Delta\tau^2])^{1/4}}\exp\left(-i\Omega_k t - \frac{t^2}{2(T^2+\Delta\tau^2)}\right),
  \en{chi}
then    from Eq.   (\ref{Eq19}) we get 
\be
p(\bt) = \prod_{k=1}^N|\chi_k(t_k)|^2.
\en{pbtnew}
Now we can recast the probability density of Eq. (\ref{Eq15})   in the following form
\begin{eqnarray}
 \label{Ad1}
&& \!\!\!\!\!\!\!  p^{(b)}_\bl(\bt)\! = \!  \frac{1}{N!} \sum_{\sigma_1,\sigma_2}\widetilde{\cJ}(\bt;\sigma_1\sigma^{-1}_2) \prod_{k=1}^N \widetilde{\mathcal{U}}^*_{\sigma_1(k),l_k}(t_k)\widetilde{\mathcal{U}}_{\sigma_2(k),l_k}(t_k),  \nonumber\\
 && \!\!\!\!\!\!\! \widetilde{\mathcal{U}}_{k,l}(t)  \equiv   U_{k,l}\chi_{k}(t).
  \end{eqnarray} 
For  $\Delta\tau=0$, i.e.,  photons in pure states,  we get from Eq. (\ref{Eq20}) $\widetilde{\cJ}(\bt;\sigma)=1$ and  Eq. (\ref{Ad1}) reduces to    the single permanent expression of Eq. (\ref{ppure}) (in this case $\chi_k(t) = \phi_{k,0}(t)$). In any experiment, however, mixed states of photons invariably lead to partial distinguishability with some nontrivial distinguishability function   $\widetilde{\cJ}(\bt;\sigma) $ in  Eq. (\ref{Ad1}), for  our Gaussian model given by Eq. (\ref{Eq20}).   

We already know from Eq.~(\ref{Eq17}) that   the proper  distinguishability function $\widetilde{\cJ}(\bt;\sigma)$  averaged over  $p(\bt)$    coincides with $J(\sigma)$.  There is, moreover, an  insightful similarity   relation,  in the functional dependence on $\sigma$, between  the proper  distinguishability functions of the two setups of figure \ref{F1} for small $\eta\ll 1$.  Namely, the argument in the exponent in the case of $\widetilde{\cJ}(\bt;\sigma)$ Eq. (\ref{Eq20}),  averaged over random photodetection times $\bt$ with the probability density $p(\bt)$ Eq. (\ref{pbtnew}),  gives    $-\eta^2(N-C_1(\sigma))$, i.e.,   the argument in the exponent of $J(\sigma)$   in Eq. (\ref{JA}). Indeed,  there are exactly $N-C_1(\sigma)$ nonzero average squared  differences $ \overline{(t_k - t_{\sigma(k)})^2}$ with the same value  
 \[
  \int\rd t_1\int \rd t_2 |\chi(t_1)|^2|\chi(t_2)|^2 (t_1-t_2)^2 = T^2 + \Delta\tau^2. 
 \]
 The similarity  of   partial distinguishability of photons  in the two setups reveals    that    distinguishability due to mixed states  is not  compensated for by increasing  photodetection    time resolution    up to arbitrarily  sharp  resolution.

 Note that   the output probability formula  given by  Eqs. (\ref{chi}) and (\ref{Ad1})  corresponds   a  non-unitary spatiotemporal transformation matrix $\widetilde{\mathcal{U}}_{k,l}(t_j)$  for any finite photon  pulse width $T$. In contrast, the same probability density in the form of Eq. (\ref{Eq15}) has the desired unitarity feature of $\mathcal{U}_{k,l}(t_j)$, where the temporal part of the combined spatiotemporal ``interferometer" is  given by the   Fourier transform. Thus, according to Eq. (\ref{Eq15}),    photodetection with precise time of arrival resolution  also  performs a unitary  transformation, additional to  the one  performed by a spatial interferometer.  By doing this, it  converts  the running phases, $e^{-\Omega_k t}$,    of photons into operating modes, in the terminology of Refs. \cite{VS2014, PartDist},  thus they are  not part of internal states in Eq. (\ref{Eq6}).    This is reflected   in the fact  that (as we show in the next section)  while   the rescaled function   $\widetilde{\cJ}(\bt;\sigma)$ of Eq. (\ref{Eq20}) is the proper  distinguishability function,   it is nevertheless  the   function   $\cJ(\bt;\sigma)$ of Eq. (\ref{Eq18}), or, equivalently,   $\widetilde{\cJ}(\bt;\sigma)$ jointly with $p(\bt)$,  describe the quality of  multiphoton interference   in the setup of figure \ref{F1}(b).  
  
\section{Measure of indistinguishability and  quality of experimental      boson sampling      }
\label{sec3}
 
Let us now quantify how  distinguishability affects the quality of multiphoton interference and  boson sampling  in   the two setups considered in the previous section, where the quality is measured by closeness  to the ideal case with completely indistinguishable bosons.   

Previously, for the setup in figure \ref{F1}(a), we have introduced \cite{TightBound} a measure of indistinguishability  of  bosons  (describing also the quality of multiphoton interference on a spatial multiport) given by $1-d_s$, where $d_s$ is the   projection of  the  internal state of bosons on the  symmetric  subspace (which corresponds to completely indistinguishable bosons \cite{VS2014,PartDist}).  This measure serves as an  upper bound on    the   total variation distance between the respective distributions with completely indistinguishable and partially distinguishable bosons.      The  total variation distance between two distributions is equal to the largest possible difference in  probabilities that the two  distributions can assign to the same event, it was also employed in the analysis of computational  complexity of boson sampling in Ref. \cite{AA}. It was also argued in Ref. \cite{TightBound} that the bound must be  tight, since  $d_s$ is the  probability that   bosons behave as   completely indistinguishable.   Below we find an equivalent     measure  of indistinguishability for   the setup   of figure \ref{F1}(b).

Let us now    introduce the   ideal  case   (i.e., with completely indistinguishable bosons) of the two setups  of figure \ref{F1}   by  setting   $J(\sigma)=1$ in    Eq. (\ref{Eq10}) and  $\widetilde{\cJ}(\bt, \sigma)=1$ in Eq. (\ref{Ad1}).  Denote  the probabilities (probability density for setup (b)) in the ideal case by $ \mathring{p}^{(a)}_\bl$ and $ \mathring{p}^{(b)}_\bl(\bt)$, respectively.  From Eq. (\ref{Eq4}) we get  
\be
 \mathring{p}^{(a)}_\bl = \frac{1}{N!} \left| \sum_\sigma \prod_{k=1}^N U_{\sigma(k),l_k}  \right|^2 = \frac{|\mathrm{per}(U[1...N|l_1...l_N]|^2}{N!},\\
 \en{Eq21}
where $U[1...N|l_1...l_N]$ is a submatrix of $U$ on rows $1,\ldots,N$ and columns $l_1,\ldots, l_N$. Similarly,  setting $\widetilde{\cJ}(\bt;\sigma)=1$ in  Eq.  (\ref{Ad1}) we obtain
\be
 \mathring{p}^{(b)}_\bl(\bt) = \frac{|\mathrm{per}(\widetilde{\mathcal{U}}[1...N|(l_1,t_1)...(l_N,t_N)]|^2}{N!},
\en{Eq22}
where $\widetilde{\mathcal{U}}[1...N|(l_1,t_1)...(l_N,t_N)]$ is  a  similar  submatrix of   $\widetilde{\mathcal{U}}_{k,l}(t)$ of Eq. (\ref{Ad1}). 

One comment is in order on Eq. (\ref{Eq22}).  From the  previous section we know that in the case of setup (b)  photons in whatever pure states  are  completely indistinguishable.     However,    the output distribution  of Eq. (\ref{Ad1}) for photons in mixed states   can approximate that of Eq. (\ref{Eq22}) for photons in pure states only if in the latter case   the     same distribution   $p(\bt)$ of detection times appears,     Eq. (\ref{pbtnew}), since $p(\bt)$ is   independent of the proper distinguishability function $\widetilde{\cJ}$.  Therefore, in Eq. (\ref{Eq22})  we  choose    pure state  of photon $k$ to be $\chi_k(t)$  of Eq. (\ref{chi}) (with   $\Delta\tau\ne0$, in contrast to  Eq. (\ref{ppure})), which guarantees the same $p(\bt)$.  

Consider how close   the distribution of photons  at the output of  a spatial multiport  is to the ideal case in the two setups, where the closeness is given by the total variation distance:
\begin{eqnarray}
\label{Eq25}
&&\mathcal{D}^{(a)} = \frac12\sum_{l_1=1}^M\ldots \sum_{l_N=1}^M   |p^{(a)}_\bl  - \mathring{p}^{(a)}_\bl |,\\
&&\!\!\! \mathcal{D}^{(b)} = \frac12 \int\rd t_1 \ldots \int \rd t_N \sum_{l_1=1}^M\ldots \sum_{l_N=1}^M |p^{(b)}_\bl(\bt) -  \mathring{p}^{(b)}_\bl(\bt)|.\nonumber\\
\label{Eq26}
&&
\end{eqnarray}
 One comment is in order on Eqs. (\ref{Eq25})-(\ref{Eq26}). On the right hand side    we have rescaled probabilities of Eq. (\ref{Eq10}) and (\ref{Eq14}), where  output ports   $\bl$ (and also photodetection times $\bt$)       appear in the summations (integrals), instead of Fock state occupations. The output  probability of setup (a) is  symmetric in $\bl$  and that of setup (b) in $(\bl,\bt)$, thus  permutations of order in   $\bl$ (respectively, in  $(\bl,\bt)$) do  not change the  probabilities.    Since, moreover,   both distributions are  appropriately rescaled,   Eqs.  (\ref{Eq25})-(\ref{Eq26})      correctly give the    total variation distance.  For instance, in the case of setup (a)   by combining all the possible sequences of  output ports $\bl$ corresponding to the same output  configuration $\m$  and using the summation identity of Eq. (\ref{idSUM})  we get
\[
\sum_{l_1=1}^M\ldots \sum_{l_N=1}^M   |p^{(a)}_\bl  - \mathring{p}^{(a)}_\bl | = \sum_{\m}| \mathring{\tilde{p}}_\m-\tilde{p}_\m|,
\]
where $\tilde{p}_\m$ is the   probability given in Eq. (\ref{Eq4}) and $\mathring{p}^{(a)}_\m $ is the respective probability  for $J(\sigma)=1$ (the ideal case). 

In Ref. \cite{TightBound} for  identical internal states $\rho_k =\rho$  it was shown that 
\be
\mathcal{D}^{(a)} \le 1- d_s,\quad   d_s =  \mathrm{Tr}\left(S_N \rho\otimes \ldots \otimes \rho\right) = \frac{1}{N!}\sum_{\sigma}J(\sigma). 
\en{Eq27}
where $S_N$ is defined in Eq. (\ref{Eq12}). We remind here that   $d_s$ defined in Eq. (\ref{Eq27})  is  the probability that   internal state of $N$  single bosons,  $\rho\otimes \ldots \otimes \rho$,  is a state of $N$ completely indistinguishable bosons, in this case 
\be
\mathring{\rho} \equiv \frac{S_N\rho\otimes \ldots \otimes \rho S_N}{d_s},
\en{rho_ind}
 since $S_N$   is a projector on a symmetric state. Therefore, $d_s$ tells us  how indistinguishable bosons are in an experimental setup  (more details in Ref. \cite{TightBound}).   

Now let us  find an equivalent    measure,  which replaces $d_s$ of Eq. (\ref{Eq27}),  in   the case of setup of figure \ref{F1}(b).  To this goal, let us generalize  the derivation of the upper bound in Eq. (\ref{Eq27})  to setup (b). The   probability  of  Eq. (\ref{Ad1}), in comparison  to that of Eq.  (\ref{Eq10}), has one    essential  new feature: dependence of the distinguishability function $\widetilde{\cJ}$ on the output data (photodetection times $\bt$).  Notwithstanding this fact,   the   main  idea of Ref. \cite{TightBound}  applies also to setup (b): we  recast the output probability in a form where the distinguishability function serves as a  ``state"  in some auxiliary linear space spanned by $N!$ permutations, whereas  the  spatial  interferometer   $U$  and  photodetection combine to a corresponding  POVM in that linear  space.   This simple  trick not only  allows us  to  derive a bound on the total variation distance $\mathcal{D}^{(b)}$ of Eq. (\ref{Eq26}),   similar to that of Eq. (\ref{Eq27}),   but also to   find the   measure of indistinguishability  for setup (b),   equivalent to  that for setup (a).  

Let us introduce an auxiliary  linear space spanned by $N!$ basis vectors $|\sigma\rangle$, where one vector is introduced for each permutation $\sigma$ of $N$ objects.   Next, we    introduce a ``state" (more precisely, state density) $\mathbf{J}(\bt)$ corresponding to  the distinguishability function $\widetilde{\cJ}(\bt;\sigma)$: 
\be
\langle\sigma_1|\mathbf{J}(\bt)|\sigma_2\rangle \equiv  \frac{1}{N!} \widetilde{\cJ}(\bt;\sigma_1\sigma^{-1}_2).   
\en{Eq28} 
The distinguishability function $\widetilde{\cJ}(\bt;\sigma)$ Eq. (\ref{Eq20})   is  a positive-semidefinite  function  over permutations, i.e.,  for any complex-valued  function  $z(\sigma)$  we   have 
 \begin{eqnarray*}
&& \sum_{\sigma_1,\sigma_2} \widetilde{\cJ}(\bt;\sigma_1\sigma_2^{-1})z(\sigma_1) z^*(\sigma_2)  \\
&&  =  \sum_{\sigma_1,\sigma_2}  z(\sigma_1)z^*(\sigma_2)\prod_{k=1}^N\frac{\langle t_{\sigma_2(k)}|\rho|t_{\sigma_1(k)}\rangle }{   \langle t_k| \rho|t_k\rangle }\ge 0.
 \end{eqnarray*}
Therefore, the introduced operator  in Eq. (\ref{Eq28})  is a positive semi-definite operator   in the auxiliary linear space,  normalized  as follows 
 \be
 \mathrm{tr}(\mathbf{J}(\bt)) \equiv\sum_\sigma  \langle\sigma|\mathbf{J}(\bt)|\sigma\rangle = 1. 
 \en{normJhat}
 Next, we introduce   rank-1 POVM  on vectors $|Z_\bl(\bt)\rangle$ in the auxiliary linear  space, where
\be
\langle\sigma|Z_\bl(\bt)\rangle \equiv \prod_{k=1}^N \widetilde{\mathcal{U}}_{\sigma(k),l_k}(t_k),
\en{Eq29} 
with $\widetilde{\mathcal{U}}_{k,l}(t)$ from Eq. (\ref{Ad1}). 
The probability density of Eq. (\ref{Ad1}) becomes an average in the auxiliary linear space, 
\be
p^{(b)}_\bl(\bt) = \langle Z_\bl(\bt)|\mathbf{J}(\bt)|Z_\bl(\bt)\rangle.
\en{Eq30}
In the ideal case,   i.e., with $\widetilde{\cJ}(\bt;\sigma)=1$,    Eq. (\ref{Eq28}) tells us that the  corresponding auxiliary ``state" is a projector
\be
\mathring{\mathbf{J}}(\bt) = |S\rangle\langle S|, \quad \langle \sigma|S\rangle \equiv \frac{1}{\sqrt{N!}}.
\en{Eq32} 
The key observation for below derivation is that   $\mathbf{J}(\bt)$ of Eq. (\ref{Eq28})  has $|S\rangle$ of Eq. (\ref{Eq32}) as an eigenvector,  as can be easily established by   verification using the definition. 
We have therefore
\be
\mathbf{J}(\bt) = \lambda(\bt) |S\rangle\langle S| + (1-\lambda(\bt))\mathbf{J}^{(\perp)}(\bt), 
\en{Eq33}
with
\be
\lambda(\bt) = \langle S|\mathbf{J}(\bt)|S\rangle = \frac{1}{N!} \sum_\sigma \widetilde{\cJ}(\bt;\sigma)\le 1,
\en{Eq34}
where a positive semi-definite  operator $ \mathbf{J}^{(\perp)}(\bt)$ is normalized by    $ \mathrm{tr}( \mathbf{J}^{(\perp)}(\bt))=1$  and satisfies the orthogonality condition  \mbox{$\mathbf{J}^{(\perp)}(\bt)|S\rangle=\langle S| \mathbf{J}^{(\perp)}(\bt) = 0$.}

Note that  due to our specific  choice of the ideal case in Eq. (\ref{Eq22}) with $\chi_k(t)$ of Eq. (\ref{chi}),  all the introduced ``states" in the auxiliary linear space correspond to  the same probability density $p(\bt)$,    given by Eq. (\ref{pbtnew}). Indeed,  in  the ideal case, by repeating the calculation of Eq. (\ref{Eq19}) for ${p}^{(b)}_\bl(\bt)|_{\lambda(\bt)=1}=\mathring{p}^{(b)}_\bl(\bt) $, we get 
\begin{eqnarray}
\label{Eq36}
\sum_{l_1=1}^M\ldots \sum_{l_N=1}^M \mathring{p}^{(b)}_\bl(\bt)   =  \prod_{k=1}^N|\chi_k(t_k)|^2.
\end{eqnarray}
\medskip
From  Eq. (\ref{Eq33}) we   obtain the same result for the  probability density  $p^{(b)}_\bl(\bt)|_{\lambda(\bt)=0}$ corresponding to the ``state"  $\mathbf{J}^{(\perp)}(\bt)$. Hence,  the same holds for $p^{(b)}_\bl(\bt)|_{\lambda(\bt)=f(\bt)}$ with  any non-negative function $f(\bt)\le 1$, e.g.,   for the probability $p^{(b)}_\bl(\bt)|_{\lambda(\bt)=1/2}$ corresponding to the state $ \frac12\left(|S\rangle\langle S| +\mathbf{J}^{(\perp)}(\bt)\right)$. This important fact will be used below.  

 Rewriting the total variation distance of Eq. (\ref{Eq26})  in the introduced  notations, we obtain  
\begin{widetext}
\begin{eqnarray}
 \label{Eq35}
\mathcal{D}^{(b)} & = & \frac12 \int\rd t_1 \ldots \int \rd t_N \sum_{l_1=1}^M\ldots \sum_{l_N=1}^M \left|  \langle Z_\bl(\bt)|\mathring{\mathbf{J}}(\bt)-\mathbf{J}(\bt)|Z_\bl(\bt)\rangle\right| \nonumber\\
&=& \frac12 \int\rd t_1 \ldots \int \rd t_N (1-\lambda(\bt)) \sum_{l_1=1}^M\ldots \sum_{l_N=1}^M \left|  \langle Z_\bl(\bt)| \left(|S\rangle\langle S| -\mathbf{J}^{(\perp)}(\bt)\right)|Z_\bl(\bt)\rangle\right|\nonumber\\
&\le &   \int\rd t_1 \ldots \int \rd t_N (1-\lambda(\bt)) \sum_{l_1=1}^M\ldots \sum_{l_N=1}^M  \langle Z_\bl(\bt)|  \frac12\left(|S\rangle\langle S| +\mathbf{J}^{(\perp)}(\bt)\right)|Z_\bl(\bt)\rangle\nonumber\\
&=&    \int\rd t_1 \ldots \int \rd t_N (1-\lambda(\bt)) \sum_{l_1=1}^M\ldots \sum_{l_N=1}^M p^{(b)}_\bl(\bt)|_{\lambda(\bt)=\frac12}  = \int\rd t_1 \ldots \int \rd t_N (1-\lambda(\bt))p(\bt)\nonumber\\
&=&   1- \frac{1}{N!} \sum_\sigma \int\rd t_1 \ldots \int \rd t_N   {\cJ}(\bt;\sigma)= 1-d_s,
\end{eqnarray}
\end{widetext}
where we have used that       $ \frac12\left(|S\rangle\langle S| +\mathbf{J}^{(\perp)}(\bt)\right)$ is   positive semi-definite operator in the auxiliary linear space corresponding  to $\lambda(\bt)=1/2$ in Eq. (\ref{Eq33}),      have taken into account the above discussed fact that  $p^{(b)}_\bl(\bt)|_{\lambda(\bt)=1/2} = p(\bt)$, with  $p(\bt)$ of  Eq. (\ref{pbtnew}), used the relations between the distinguishability functions $\cJ$, $\widetilde{\cJ}$, and $J$, given by Eqs. (\ref{Eq17}) and (\ref{pbtnew}), and the definition of $d_s$ in Eq. (\ref{Eq27}). 

Thus,  for the  setup in figure \ref{F1}(b) we have obtained the same   bound on the total variation distance to the ideal case as for the  setup in figure  \ref{F1}(a). The  physical reason for this  is that in these two  setups the (total) probability that the input  photons  are  completely indistinguishable   is \textit{the same}.   Let us show that   $d_s$ of Eq. (\ref{Eq27})   is  also the total  probability that photons are completely indistinguishable in the case of  setup of figure \ref{F1}(b).  
Indeed,  given  photodetection   times $\bt$,   by the fact that the   projector $|S\rangle\langle S|$ of Eq. (\ref{Eq33}) corresponds to  the  ideal case in setup (b),  the   conditional on $\bt$   probability density that  photons  in setup (b) are completely indistinguishable  reads $d^{(b)}_s(\bt)= \lambda(\bt)$, where $\lambda(\bt)$ is the eigenvalue   of $\mathbf{J}(\bt)$  in  Eq (\ref{Eq33}). This can be also established using     the internal state, conditional on  given photodetection times $\bt$, which is easily  obtained  by comparing Eqs. (\ref{Eq5}) and   Eq. (\ref{rho_ind}) with Eq.  (\ref{Eq16})
 \begin{eqnarray}
\label{dsbt}
&&d^{(b)}_s(\bt)\equiv  \frac{\left[\prod_{k=1}^N{}^{\otimes} \langle t_k|\right] S_N   \rho \otimes \ldots \otimes \rho  S_N \left[\prod_{k=1}^N{}^{\otimes}|t_k\rangle\right]}{p(\bt)}\nonumber\\
&&  =   \frac{\mathrm{Tr}\left( S_N |t_1\rangle\langle t_1| \otimes \ldots \otimes  |t_N\rangle\langle t_N| S_N \rho \otimes \ldots \otimes   \rho   \right)}{p(\bt)} \nonumber\\
&& = \frac{1}{N!}\sum_\sigma \frac{\cJ(\bt;\sigma)}{p(\bt)} =    \frac{1}{N!}\sum_\sigma \widetilde{\cJ}(\bt;\sigma) = \lambda(\bt),
\end{eqnarray}
where we have used  the definition of $S_N$    (\ref{Eq12}) and  Eq.   (\ref{Eq34}).   The total probability is therefore 
 \begin{eqnarray}
 \label{Eq37}
 && d^{(b)}_s \equiv  \int\rd t_1 \ldots \int \rd t_N p(\bt) d^{(b)}_s(\bt) \nonumber\\
 && = \frac{1}{N!}\sum_\sigma  \int\rd t_1 \ldots \int \rd t_N  \cJ(\bt;\sigma) =   d_s,
\end{eqnarray}
by Eqs. (\ref{Eq34}) and (\ref{dsbt}), the relation between the distinguishability functions in Eq. (\ref{Eq17}) and the definition of $d_s$ (\ref{Eq27}). Hence, the bound $1-d_s$   in Eqs. (\ref{Eq27}) and (\ref{Eq35}) bears   the same  physical  interpretation  for both setups of    figure \ref{F1}.

As an application, consider how small the   purity of single photons should be for a good quality experimental boson sampling?  Our  common upper  bound $1-d_s$ can  give a sufficient purity.  An explicit expression     can be obtained with the use of the approximation of  Eq. (\ref{JA}), which is very good  for  $\eta \lesssim 0.125$   (see figure \ref{F2} in appendix \ref{appA}). We have (see mathematical details in appendix \ref{appB})
\begin{eqnarray}
\label{Eqd_s}
&& d_s \approx  \exp\left(-\eta^2 N  \right)\sum_{n=0}^N \frac{1}{n!}\left( e^{\eta^2} - 1\right)^n  \nonumber\\
&&=  \exp\left(-\eta^2 N  \right)\left( e^{\eta^2} +  \sum_{n=2}^N \frac{1}{n!}\left( e^{\eta^2} - 1\right)^n\right).
\end{eqnarray}
 For a given  purity of single photons, $\mathcal{P} \equiv \mathrm{Tr}( \rho^2) \approx e^{-2\eta^2} $ by Eq. (\ref{Eq8A}),  from Eq. (\ref{Eqd_s})    we get
\be
\mathcal{D}^{(a,b)} \le 1 -  d_s \approx  \mathcal{P}^{\frac{N-1}{2}}\left(  1 +  \sum_{n=2}^N \frac{ \left(1-\sqrt{\mathcal{P}} \right)^n}{n! \mathcal{P}^{\frac{n-1}{2}}}\right).
\en{EqBound}
For illustration, to have   up to  $10\%$ deviation from the ideal case of $N$-photon experimental boson sampling for  $20 \le N \le  50$ photons our estimate necessitates the  photon state  purity    $0.989 \le \mathcal{P}\le  0.996$.

\section{Conclusion}
\label{sec4}
We have considered how  inevitable  noise  affects multiphoton interference with  single photons in two different setups: in the   setup  with photons of the same  central frequency on a spatial interferometer and  slow detectors,  incapable of time  resolution,  proposed for realizing  boson sampling  \cite{AA},   and  in the setup with photons of different  central frequencies on the same spatial interferometer and  fast detectors, capable of precise time resolution,    the so-called ``multiboson correlation sampling"  of  Ref. \cite{TL}. To be able to   carry out all calculations explicitly, we have focused on  the specific model of   mixed-state  input photons  having Gaussian temporal shapes  with  random arrival times    governed  by  a Gaussian distribution.  We have revealed  clear physical interpretations of main results, thus the conclusions are independent of the  considered  model.
 
We have found that  the  partial distinguishability  theory of Refs.   \cite{VS2014,PartDist} applies also to  multiphoton interference  with photodetection capable of  precise time resolution,  moreover,  the output probability distribution  is given by a similar mathematical expression  as for  the time-unresolved photodetection, where a function on a permutation group describes partial distinguishability of photons.     For input photons  in   pure   states,  the  output probability is given by the absolute value squared of a single matrix permanent,   reproducing  the results of  Refs. \cite{TL1,TL}.  In this case photons are   completely indistinguishable, since   photodetection with precise time resolution turns the otherwise  internal (temporal) states of photons  into the operating modes by coherently mixing different paths of photons.   Separation of  degrees of freedom into  the operating modes and the internal states is the  prerequisite  for application of the general theory of Refs. \cite{VS2014,PartDist}.  We find that, while the average pure states of photons are operating modes under the photodetection with precise photon state resolution, the fluctuations about them are not,  thus photons in mixed states  are always partially distinguishable.

 Whereas,  with  the   operating modes and  internal states being properly identified,  the  applicability of the theory of Refs. \cite{VS2014,PartDist}  to    photodetection with  precise time resolution  may not come as   big surprise (though other views exist \cite{TL1}),   the  striking similarity of    distinguishability   of photons due to mixed states  under    photodetection with  and without precise   time resolution   and the same  probability of photons to behave as  completely indistinguishable in the two cases are  rather  surprising. These facts have broad implications. For instance,   purity of photons, usually reported alongside with photon  distinguishability in current experiments, e.g., Refs. \cite{SP1,SP2},   is in fact, the unavoidable distinguishability, whatever  the photodetection scheme employed.    This implies that there is no advantage in quality  of    the so-called ``multiboson correlation sampling"  of  Ref. \cite{TL} over the standard boson sampling  \cite{AA} if  photons of the same purity are used in both schemes.

Though we have focused  only on   state-resolving photodetection   where  times of arrival  are resolved by detectors,    there is nothing special about the particular detection scheme considered. Our  main results   on distinguishability due to mixed states (noise)  in an experimental setup are also applicable    to frequency-resolved photodetection of photons with different times of arrival, as for instance, in a recent experiment \cite{FreqRes},  with  the roles played by time and frequency   interchanged. 

Finally, since  the    theory of Refs.  \cite{VS2014,PartDist} is     applicable to  the case of    photodetection  with precise time (or frequency) resolution,   all    the  previous  results  apply also to such setups as well, e.g., the  fundamental limits on  the quality of boson sampling experiments \cite{VS2014,TightBound} and the  classical simulation algorithm, due to partial distinguishability \cite{JR}.    We believe that this general conclusion provides a basis     for    assessment of      experimental  boson sampling  setups with different detection schemes  for demonstration of   quantum advantage over classical computers.

 \medskip
\section{acknowledgements}
V. S.    was supported by the National Council for Scientific and Technological Development (CNPq) of Brazil,  grant 307813/2019-3  and by the S\~ao Paulo Research Foundation (FAPESP), grant 2018/24664-9.  M. E. O. B. was supported  by the Coordenação de Aperfeiçoamento de Pessoal de Nível Superior   (CAPES) of Brazil,  Finance Code 001.

\appendix
\section{ On derivation  of Eq. (\ref{Eq8}) }
\label{appA}
Here we derive the expression for $  \mathrm{Tr}\left(\rho^n\right)$ of Eq. (\ref{Eq8}). We have (with index  $j$  mod $n$)
 \begin{widetext}
 \begin{eqnarray}
 \label{Aq0}
  \mathrm{Tr}\left(\rho^n\right) &=& \int\rd \tau_1p(\tau_1)\ldots \int \rd\tau_np(\tau_n) \prod_{j=1}^n \langle\phi_{j,\tau_j}|\phi_{j+1,\tau_{j+1}}\rangle \nonumber\\
  &=& \frac{1}{(\pi \Delta\tau^2)^{n/2}}\int\rd \tau_1\ldots \int \rd\tau_n \exp\left( - \frac{\sum_{j=1}^n \tau_j^2}{\Delta\tau^2}-\frac{\sum_{j=1}^n (\tau_j-\tau_{j+1})^2}{4T^2}\right)\nonumber\\
  &=& \frac{1}{\pi ^{n/2}}\int\rd x_1\ldots \int \rd x_n \exp\left( - \sum_{i,j=1}^n x_i A_{ij}x_j\right) = \frac{1}{\sqrt{\mathrm{det} A}},
  \end{eqnarray}
\end{widetext}
where  $x_j\equiv \frac{\tau_j}{\Delta\tau}$ and $A$ is  the circulant matrix,   with the only  nonzero elements $A_{jj} =  1 + 2\eta^2$ and \mbox{$A_{j,j\pm 1} = -\eta^2$}. Determinant of $A$ is given by an   explicit  analytical expression \cite{Detcirculant}.  For a general   $n$-dimensional circulant matrix  
\be
A = \left(\begin{matrix} a_0 & a_{n-1} & \ldots & a_1 \\
a_1 & a_{0} & \ldots & a_2 \\
\vdots & \vdots & \ddots & \vdots \\
a_{n-1} & a_{n-2} & \ldots & a_0 \\
\end{matrix}
\right),
\en{Aq1}
denoting  $\xi \equiv e^{\frac{2i\pi}{n}}$, we get \cite{Detcirculant}
\be
\det{A} = \prod_{j=0}^{n-1} f(\xi^j),\quad f(\xi) \equiv a_0 +a_1 \xi+\ldots + a_{n-1}\xi^{n-1}.
\en{Aq2}
In case of matrix $A$ in Eq. (\ref{Aq0}) we have only three non-zero elements: $a_0 = 1+ 2\eta^2$ and $a_1 = a_{n-1} =  -\eta^2 $. Using  $\xi^{n-1} = \xi^{-1}$, we obtain  
\begin{eqnarray}
\label{Aq3}
&& \det{A} = \prod_{j=0}^{n-1}\left[ 1+ 2\eta^2 - 2\eta^2\cos\left(\frac{2\pi j}{n}\right)\right]\nonumber\\
&& = \left( 1+ 2\eta^2 \right)^n \prod_{j=0}^{n-1} \left[ 1- \frac{2\eta^2}{1+2\eta^2} \cos\left(\frac{2\pi j}{n} \right)\right]\nonumber\\
&& =  \left( 1+ 2\eta^2 \right)^n( X_+^n-X_-^n)^2,
\end{eqnarray}
where  in the last step we have used an identity of Ref. \cite{PBM} for the product and introduced  $X$ and $Y$  as follows 
\be
X^2_\pm = \frac12\left(1\pm \frac{\sqrt{1+4\eta^2}}{1+2\eta^2} \right).
\en{Aq4}
Eqs. (\ref{Aq0}) and (\ref{Aq3}) and (\ref{Aq4}) give the expression in Eq. (\ref{Eq8}) of the main text. 

\begin{figure}[htb]
\begin{center}
     \includegraphics[width=.315\textwidth]{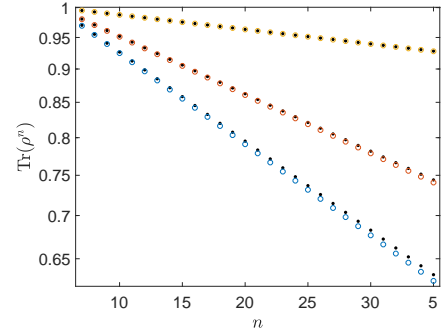}
     \caption{Higher-order purity $ \mathrm{Tr}\left(\rho^n\right)$ (blobs) and its approximation by the  exponent in Eq. (\ref{Eq8})   (circles)  vs. $n$  for several values  of  $\eta=\frac{\Delta\tau}{2T}$, from top to bottom: $\eta = 0.05; 0.1; 0.125$.  \label{F2} }
   \end{center}
\end{figure}

Finally, to judge how good is the exponential  approximation  in Eq. (\ref{Eq8A}) to  the exact expression for  $\mathrm{Tr}\left(\rho^n\right)$,   Eq. (\ref{Eq8}) in  the main text, a numerical comparison of the two  is given in Fig. \ref{F2}.

\section{ On derivation  of Eq. (\ref{Eqd_s}) }
\label{appB}

We can use the method of generating function   to compute the cycle sum $Z_N$ in 
\be
d_s \approx \frac{1}{N!}\sum_{\sigma} \exp\left(- \eta^2 [N-C_1(\sigma)]\right) =  \exp\left(- \eta^2 N\right)Z_N,
\en{Bq1}
with $Z_N \equiv \frac{1}{N!}\sum_\sigma \zeta^{C_1(\sigma)}$  and $\zeta\equiv e^{ \eta^2 }$. Using that $\sum_{n=1}^N n C_n(\sigma) = N$ and \cite{Stanley}
\[
\frac{1}{N!}\sum_\sigma \left(\ldots \right) =  \sum_{C_1,\ldots,C_N} \frac{\left(\ldots\right)}{\prod_{n=1}^N n^{C_n} C_n!} 
\]
we get the  generating function $F(X) \equiv  \sum_{N\ge1} Z_N X^N$ as follows 
\begin{eqnarray}
\label{Bq2}
&& F(X)  = \sum_{N\ge 1} \frac{1}{N!}\sum_\sigma  (X\zeta)^{C_1(\sigma)} \prod_{n=2}^N X^{nC_n(\sigma)}  \nonumber\\
&& = \sum_{C_1,\ldots,C_N} \frac{  (X\zeta)^{C_1} \prod_{n=2}^N X^{nC_n}}{\prod_{n=1}^N n^{C_n} C_n!} \nonumber\\
&& = \exp\left([\zeta-1]X +\sum_{n=1}^\infty \frac{X^n}{n} \right)\nonumber\\
&&= \frac{ \exp\left([\zeta-1]X \right)}{1-X}.
\end{eqnarray}
Therefore
\be
Z_N = \left. \frac{1}{N!} \frac{d^N F(X)}{dX^N}\right|_{X=0}  = \sum_{n=0}^N \frac{(\zeta-1)^n}{n!},
\en{Bq3}
which results in the expression for $d_s$ in   Eq. (\ref{Eqd_s}) of section \ref{sec3}.


\bigskip
\begin{thebibliography}{99}

\bibitem{Reviewbs}  D. J. Brod, E. F. Galv\~ao; A. Crespi,  R. Osellame,  N.  Spagnolo and F. Sciarrino,  Advanced Photonics, \textbf{1}, 034001 (2019). 

\bibitem{AA} S. Aaronson and A. Arkhipov,  	arXiv:1011.3245 [quant-ph]; Theory of Computing \textbf{9},  143 (2013).
  
\bibitem{20ph60mod} Hui Wang \textit{et al}, Phys. Rev. Lett. \textbf{123,}  250503 (2019). 

\bibitem{HOM} C. K. Hong, Z. Y. Ou, and L. Mandel, Phys. Rev. Lett. \textbf{59},   2044 (1987).


\bibitem{R1} P. P. Rohde and T. C. Ralph, Phys. Rev. A \textbf{85}, 022332  (2012).

 \bibitem{VS2014} V. S. Shchesnovich,   Phys. Rev. A \textbf{89}, 022333 (2014).

\bibitem{TightBound} V. S. Shchesnovich, 	 Phys. Rev. A   \textbf{91}, 063842
4  (2015). 


\bibitem{Jelmer2018}   J. J. Renema, A. Menssen, W. R. Clements, G. Triginer, W. S. Kolthammer, and I. A. Walmsley,
 Phys. Rev. Lett. \textbf{120}, 220502  (2018). 
 
\bibitem{KK}  G. Kalai and G. Kindler,  	arXiv:1409.3093 [quant-ph]. 

\bibitem{LP} A. Leverrier and R. Garc{\'i}a-Patr{\'o}n,  arXiv:1309.4687 [quant-ph].

\bibitem{A}  A. Arkhipov,  Phys. Rev.  A  \textbf{92}, 062326 (2015). 

\bibitem{AB} S. Aaronson and D. J. Brod,   Phys. Rev.  A  \textbf{93}, 012335 (2016).

\bibitem{PRS} R. Garc\'ia-Patr\'on, J. J. Renema, and V. S. Shchesnovich, 		Quantum \textbf{3}, 169 (2019).  

\bibitem{OB} M. Oszmaniec and D. J. Brod, New J. Phys. \textbf{20}, 092002 (2018). 


\bibitem{VS2019} V. S. Shchesnovich, 	Phys. Rev. A \textbf{100}, 012340  (2019). 



\bibitem{MPBF} M. C. Tichy, M. Tiersch, F. Mintert, and A. Buchleitner, New Journal of Phys.  \textbf{14},  093015 (2012).

\bibitem{Exp1} Y.-S. Ra, M. C. Tichy, H.-T. Lim, O. Kwon, F. Mintert, A. Buchleitner, and Y.-H. Kim,  PNAS \textbf{110,} 1227 (2013). 


\bibitem{Rohde}  P. P. Rohde, Phys. Rev. A \textbf{91}, 012307 (2015).

\bibitem{PartDist} V. S. Shchesnovich, 	 Phys. Rev. A   \textbf{91}, 013844  (2015). 

\bibitem{Tichy} M. C. Tichy, Phys. Rev. A  \textbf{91}, 022316 (2015).


\bibitem{TL1} V. Tamma and S.   Laibacher,  Phys. Rev. Lett. \textbf{114,}  243601 (2015). 


\bibitem{SUN} M. Tillmann, S.-H. Tan, S. E. Stoeckl, B. C. Sanders, H. de Guise, R. Heilmann, S. Nolte, A. Szameit, and  P.~Walther, Phys. Rev. X \textbf{5}, 041015 (2015).


\bibitem{TL} V. Tamma and S.   Laibacher, Quantum Inf. Process \textbf{15,} 1241  (2016). 


\bibitem{MCMS}  J.-D. Urbina, J. Kuipers, S. Matsumoto, Q. Hummel, and K. Richter, Phys. Rev. Lett. \textbf{116}, 100401 (2016).

\bibitem{CrSp} M. Walschaers, J. Kuipers, and A. Buchleitner, Phys. Rev.  A \textbf{94}, 020104(R) (2016).


\bibitem{Exp2} A. J. Menssen, A. E. Jones, B. J. Metcalf, M. C. Tichy, S. Barz, W. S. Kolthammer, and I. A. Walmsley,  Phys. Rev. Lett. \textbf{118,} 153603 (2017).


\bibitem{VSMB} V. S. Shchesnovich and M. E. O. Bezerra, Phys. Rev. A \textbf{98,} 033805 (2018). 

\bibitem{DD} S. Stanisic and P. S. Turner, Phys. Rev. A \textbf{98,} 043839 (2018). 

\bibitem{QSPD} A. E. Moylett and P. S. Turner, Phys. Rev. A \textbf{97,} 062329 (2018). 


\bibitem{SPnp}  M.E. Reimer and C. Cher,  Nat. Photonics \textbf{13,}  734  (2019). 


\bibitem{SPDCtjitt}  W. P. Grice and I. A. Walmsley, Phys. Rev. A \textbf{56,}  1627 (1997). 

\bibitem{HervsPurity} E.  Meyer-Scott, N. Montaut, J. Tiedau, L. Sansoni,  H. Herrmann, T. J. Bartley, and C. Silberhorn, Phys. Rev. A \textbf{95,} 061803(R)  (2017). 

\bibitem{QDtjitt} E. B. Flagg,  S. V. Polyakov,  T.  Thomay, and  G. S. Solomon,   Phys. Rev. Lett. \textbf{109,} 163601 (2012).


\bibitem{HeraldSP} D. B. Horoshko, S. De Bi\'evre, G. Patera, and M. I.  Kolobov, Phys. Rev. A \textbf{100,} 053831  (2019). 


\bibitem{Rdetec1} S. M. Barnett, L. S. Phillips, and D. T. Pegg,   Opt. Commun. \textbf{158,} 45 (1998).

\bibitem{Rdetec2}   M. Ramilli, A. Allevi, V. Chmill, M. Bondani, M. Caccia, A. Andreoni,   J. Opt. Soc. Am. B \textbf{27} 852 (2010).


\bibitem{SP1} V. Ansari \textit{et al}, Optics Express \textbf{26,}    2764  (2018). 

\bibitem{SP2} L.  Dusanowski, S.-H.  Kwon, C.  Schneider, and S. H\"ofling,  Phys. Rev. Lett. \textbf{122,} 173602 (2019). 



\bibitem{QSBS} A. Neville, C. Sparrow, R. Clifford, E. Johnston, P. M. Birchall, A. Montanaro, A. Laing,
Nature Physics \textbf{ 13}, 1153 (2017).
 
 
 \bibitem{Gphotons} P. P. Rohde, T. C. Ralph,  and  M. A. Nielsen, Phys. Rev. A \textbf{72}, 052332 (2005).

\bibitem{Stanley}   R. P. Stanley, \textit{Enumerative Combinatorics}, 2nd ed., Vol. 1 (Cambridge University Press, 2011).


\bibitem{ContFields} K. J. Blow, R. Loudon,  S. J. D. Phoenix, and T. J. Shepherd,  Phys. Rev. A \textbf{42,}  4102 (1990). 





\bibitem{FreqRes} V. V. Orre, E.  A. Goldschmidt, A. Deshpande, A.V. Gorshkov, V. Tamma, M. Hafezi, and S. Mittal, 
Phys. Rev. Lett. \textbf{123,} 123603 (2019). 
 

\bibitem{JR}   J. J. Renema, A. Menssen, W. R. Clements, G. Triginer, W. S. Kolthammer, and I. A. Walmsley,
 Phys. Rev. Lett. \textbf{120}, 220502  (2018). 
 
 \bibitem{Detcirculant} P. J. Davis,   \textit{Circulant Matrices}  (American Mathematical Society; 2 edition   2012). 


\bibitem{PBM}  A. P. Prudnikov, Yu. A. Brychkov, and  O. I. Marichev,  \textit{Integraly i Ryady. Elementarnue  funktzii } (in Russian)  (Moscow, FizMatLit 1981).
 
 
\end{thebibliography}
\end{document}